\documentclass[aps,prb,showpacs,twocolumn]{revtex4}
\usepackage{epsfig}
\usepackage{amsmath}

\begin{document}

\title{Driving a first order quantum phase transition by coupling a
quantum dot to a 1D charge density wave}
\author{Y. Weiss, M. Goldstein and R. Berkovits}
\affiliation{The Minerva Center, Department of Physics, Bar-Ilan University,
  Ramat-Gan 52900, Israel}

\begin{abstract}

The ground state properties of a one-dimensional system with particle-hole 
symmetry, consisting of a gate controlled dot coupled to an interacting 
reservoir, are explored using the numerical DMRG method.
It was previously shown that the system's thermodynamic properties 
as a function of the gate voltage in the Luttinger liquid phase are 
qualitatively similar to the behavior of a non-interacting wire with an 
effective (renormalized) dot-lead coupling.
Here we examine the thermodynamic properties of the wire in the charge 
density wave phase, and show that these properties behave quite differently. 
The number of electrons in the system remains constant as a function of the 
gate voltage, while the total energy becomes linear. Moreover, by tuning 
the gate voltage on the dot in the charge density wave phase it is 
possible to drive the wire through a first order quantum phase transition 
in which the population of each site in the wire is inverted.

\end{abstract}


\maketitle

The properties of one dimensional (1D) interacting systems 
have attracted much interest going back half a century\cite{ll,voit}. Much
recent effort has concentrated on understanding the conductivity
and I-V characteristics of a Luttinger liquid (LL) coupled
to an impurity \cite{kane92}. These properties are probed
experimentally by measurements of the temperature dependent 
conduction through 1D systems\cite{cond}, 
and tunneling spectroscopy into 1D wires\cite{auslander05}.
The measurements of Ref. \onlinecite{auslander05} also
indicate a localization transition in the wire for low densities which
might be associated with a charge density wave (CDW).
Recently it has been realized that the occupation of an
impurity level or a quantum dot 
coupled to an interacting 1D sample may be used as a probe to its properties 
\cite{matveev,our_prev}. The occupation is experimentally accessible, for example
by the charging effect of the impurity on a quantum point contact in its
vicinity\cite{johnson04}.
A generic model for such a situation is a dot
coupled to a lead. If such a dot is controlled by a gate
one can change its orbital energy. In this paper we shall
demonstrate that by following the occupation of the dot as a function
of the gate voltage it is possible to identify a transition in the
wire from a LL to a CDW phase. 
In the CDW regime of the wire the occupation of the dot switches abruptly
once the dot orbital crosses the Fermi energy.
Moreover, by changing the gate voltage of the quantum dot
the wire is driven through a first
order quantum phase transition (QPT), in which the ground state of 
the entire system switches.

The difference between measuring the conductivity through the dot-lead system
or the local density of states at the impurity \cite{grishin04},
and probing the dot occupation using e.g. a quantum point contact must be emphasized.
It is commonly known that the special behavior of the 
LL spectrum near the Fermi energy causes the conductivity to vanish, 
even in the presence of a single impurity \cite{nazarov}. Thus measuring a 
conductivity of a system containing a LL lead and a dot is not
expected to show any level broadening at zero temperature.

Nevertheless, connecting the dot to the LL lead, while measuring
its occupation using a quantum point contact, results in a different situation.
In this case the LL acts as a reservoir for the dot, and a level
broadening can be seen. This difference results from the fact that
such a level is coupled to all the lead states, even those 
far from the Fermi energy, and thus the special behavior of the LL near $E_F$
is not expected to dominate. It was indeed shown, that
the dot's level width follows the regular Breit-Wigner
line shape, even in the presence of interactions,
although the strength of coupling between the dot
and the 1D wire is renormalized as a function of the 
interactions in the wire \cite{our_prev}.

In this paper we show that a qualitatively different behavior of the dot's 
population is exhibited once the wire is in the CDW phase.
It is known that a one-dimensional system of spinless electrons 
with nearest neighbor repulsive interactions undergoes a QPT between a LL 
phase for weak interactions, and a CDW phase for strong 
interactions \cite{luther_peschel,haldane_80,schulz}. 
The different behavior of the dot population between the LL and the CDW phases
stems from the difference between the ground states of the two phases in the wire.
It is argued that while the LL is only locally influenced 
by the coupling to the dot, a semi-infinite CDW coupled to the dot 
undergoes a QPT when the orbital's level crosses the wire's chemical potential. 
We show that a sharp jump in the dot's population occurring at this point
is accompanied by an abrupt switch of the ground state of the entire system.
This switch is not a simple level crossing, but rather a true 
QPT since it becomes a sharp transition only once the wire is semi-infinite \cite{qpt_book}.
Other hallmarks of a first order QPT, such as
a discontinuity in the first derivative of the grand potential,
as well as an inversion of the CDW order parameter as a result of the
inversion of the wire's occupation for each site, are seen.

The Hamiltonian describing a spinless fermionic system
composed of a single orbital ("dot") coupled to a 1D
nearest neighbor interacting wire ("lead"),
is given by $\hat H = \hat H_{dot} + \hat H_{dot-lead} + \hat H_{lead}$,
where
\begin{eqnarray} \label{eqn:H_dot}
{\hat H_{dot}} &=& \epsilon_{0}{\hat a}^{\dagger}{\hat a}, \\ \nonumber
{\hat H_{dot-lead}} &=& -V ({\hat a}^{\dagger}{\hat c}_{1} + 
{\hat c}^{\dagger}_{1}{\hat a}), \\ \nonumber
{\hat H_{lead}} &=& -t
\displaystyle \sum_{j=1}^{L-1}({\hat c}^{\dagger}_{j}{\hat c}_{j+1} + h.c.) \\ \nonumber
&+& I \displaystyle \sum_{j=1}^{L-1}({\hat c}^{\dagger}_{j}{\hat c}_{j} - \frac{1}{2})
({\hat c}^{\dagger}_{j+1}{\hat c}_{j+1} - \frac{1}{2}).
\end{eqnarray}
We denote the dot's energy level by $\epsilon_0$, 
$V$ ($t$) is the dot-lead (lead) hopping matrix element, and
$I$ is the nearest-neighbor interaction strength
in the lead. ${\hat a}^{\dagger}$ (${\hat a}$) is the creation (annihilation)
operator of an electron in the dot, 
and ${\hat c}_j^{\dagger}$ (${\hat c}_j$) is the creation (annihilation)
operator of an electron at site $j$ in the lead.
In the interaction term, a positive background is assumed.
The lead hopping matrix element, $t$, is taken as $1$, 
in order to set the energy scale.

In order not to increase the number of free parameters, the model discussed here 
does not contain a dot-lead interaction term. Nevertheless, we have checked that 
such a term does not influence qualitatively the results of this paper. 
We refer the reader to a previous work \cite{our_prev}, in which it was 
explicitly shown that an addition of such an interaction 
term influences only the effective position and width of the dot level.

As was previously mentioned, while the
lead is in the LL phase, the Breit-Wigner
formula with an effective dot-lead coupling describes the
dot population $n(\epsilon_0)$ quite well \cite{our_prev}. 
For a dot coupled to a LL lead in which the interaction parameter 
$g<1/2$, a jump in the dot population was 
predicted by Furusaki and Matveev \cite{matveev}. However, for the 
spinless model with nearest neighbor interaction the LL parameter is
always in the range of $1/2 \le g \le 1$,
and the Furusaki-Matveev jump is not expected.

The grand potential $\hat \Omega={\hat H}-\mu {\hat N}$ 
(${\hat N}$ is the particle number operator)
was diagonalized using a finite-size DMRG calculation 
\cite{white93,our_prev}
for different values of $V$ and $I$, with a lead of up to $L=300$ sites.
In order to obtain particle-hole symmetry, for which the LL to CDW transition
was widely discussed \cite{luther_peschel,haldane_80,schulz},
$\mu$ is set to zero. 
The total number of particles in the system is not fixed during the renormalization
process, so that the results describe the experimental situation of a 
finite section of a 1D wire which is coupled to a dot and to an external electron reservoir.
In particular, the total occupation obtained for such a system can be non-integral.
As a function of the dot's level ($\epsilon_0$), 
the following ground state properties were calculated:
the system's grand potential $\Omega$, the total number of electrons $N$
and the dot population $n_{dot}$. The population of
lead sites were also calculated 
in order to differentiate between local effects of the dot population and
global phenomena in the lead.

\begin{figure}[htb]\centering
\vskip -2.0truecm
\epsfxsize6.0cm
\epsfbox{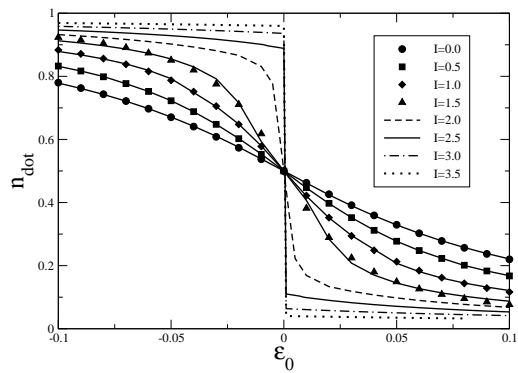}
\vskip -.5truecm
\caption{\label{fig:n_dot}
The dot population $n_{dot}$ as a function of the level energy $\epsilon_0$.
For $I<2$ (LL phase) the curves fit the non-interacting 
formula with an effective coupling 
constant $V_{eff}$ (lines - DMRG results, symbols - fit to Eq.~\ref{eqn:n_dot_exact}), 
while for $I>2$ (CDW phase) the width is zero.
Exactly at the transition point ($I=2$) $n_{dot}$ is continuous
but doesn't fit the non-interacting formula.}
\end{figure}

We begin by comparing the behavior of $n_{dot}$ in the 
two phases of the interacting lead. 
For spinless fermions with nearest-neighbor interactions, 
it is known that the system experiences
a phase transition (of a Kosterlitz-Thouless type \cite{kosterlitz}),
between the LL and the CDW phases, at $I=2t$. Indeed, the results
presented in Fig.~\ref{fig:n_dot} show a qualitative difference
between $n_{dot}(\epsilon_0)$ in these two regimes. In the LL phase
the curves fit quite well the Breit-Wigner formula \cite{our_prev}
with an effective coupling $V_{eff}$
\begin{eqnarray} \label{eqn:n_dot_exact}
{n}_{dot}(\mu,\epsilon_0) = \\ \nonumber
\frac{1}{\pi} \displaystyle \int_{-2t}^{\mu} & {\frac {\frac{V_{eff}^2}{t}
\sqrt {1-\frac{\epsilon^2}{4t^2}} }
{\frac{V_{eff}^4}{t^2} (1-\frac{\epsilon^2}{4t^2}) +
((1-\frac{V_{eff}^2}{2t^2})\epsilon - \epsilon_0)^2} d\epsilon}, 
\end{eqnarray}
while in the CDW phase the width becomes zero, and a jump in $n_{dot}$ occurs
at $\epsilon_0=0$. This jump is associated with the
degeneracy of the CDW ground state.

In order to support this argument, we first calculate the
properties of the ground state for a lead not coupled to a dot, 
by taking $V=t$ and the limit $\epsilon_0 \rightarrow 0$.
The CDW order parameter can be defined as \cite{pang_93}
\begin{eqnarray} \label{eqn:P_cdw}
{P(i) = \frac {1}{2} (-1)^i [ 2n(i)-n(i-1) - n(i+1)]},
\end{eqnarray}
where $n(i)= <{\hat c}^{\dagger}_{i}{\hat c}_{i}>$ 
is the occupation of the i'th lead site in the ground state.
The value of $P(i)$ does not change much as a function of the spatial 
coordinate, except in the vicinity of the lead's edges. 
We thus define $P=P(L/2)$.

\begin{figure}[htb]\centering
\vskip -2.0truecm
\epsfxsize5.5cm
\epsfbox{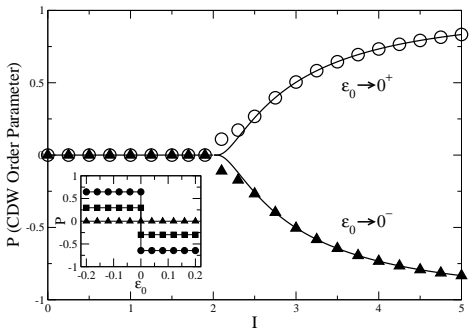}
\vskip -.5truecm
\caption{\label{fig:cdw_order}
The CDW order parameter $P$ as a function of the interaction $I$
for a 300-sites lead (DMRG - symbols, theory - lines). 
For the CDW phase $P$ is inverted
between the cases $\epsilon_0 \rightarrow 0^+$ and $\epsilon_0 \rightarrow 0^-$,
while in the LL phase both cases result in $P=0$.
Inset: $P$ as a function of $\epsilon_0$ for
$I=1.5$ (triangles), $2.5$ (squares) and $3.5$ (circles). The LL
case results in a constant $P=0$, while an inversion of 
$P$ occurs at $\epsilon_0=0$ for the two CDW cases.
}
\end{figure}

In Fig.~\ref{fig:cdw_order} the dependence of $P$ 
on the interaction strength $I$, as obtained by the DMRG method, is shown, and 
compared to the exact results \cite{baxter}. 
For $I<2$, the system is in the LL phase and indeed $P=0 \pm 10^{-4}$.
In this case the population of each lead site is $1/2$ 
and the lead is half filled; i.e. $N=L/2$.

As expected, the metal-insulator transition occurs at $I=2$.
For $I>2$, in which the system is in the CDW phase, 
the value of $P$ is finite. For values of $I$ which are far 
from the transition point ($I > 2.5$), we get a very good fit to the theory.
Near the transition point the numerics tend to emphasize the charge oscillations, 
resulting in too large values of $|P|$. This tendency, however, does not
affect the following qualitative conclusions.

As can be seen, the CDW ground state is different for the cases 
$\epsilon_0 \rightarrow 0^+$ and $\epsilon_0 \rightarrow 0^-$,
resulting in two values of $P$ (i.e., $\pm |P|$) for any value of $I$ in the CDW regime.
For $\epsilon_0 = 0$ the ground state is two-fold degenerate in the
thermodynamic limit.

In the CDW case, special care should be devoted to 
the number of electrons in the system, and
to the difference between even and odd lead length $L$. 
For $\epsilon_0 \rightarrow 0$ and $V=t$, the lead length is 
effectively $L+1$.
Denoting the ground state for $\epsilon_0 \rightarrow 0^{\pm}$  
by $\psi^{(\pm)}_0$,
and it's population in each site as $n^{(\pm)}_j$, where $j=0$ for the dot 
and $1 \le j \le L$ for
the lead sites, the particle hole symmetry implies that 
$n^{(-)}_j=1-n^{(+)}_j$ everywhere. 
For odd $L$ (even $L+1$) one has the requirement
$n^{(-)}_j=n^{(+)}_{L-j}$, and the number of electrons in both
$\psi^{(+)}_0$ and $\psi^{(-)}_0$ is $N=(L+1)/2$.
When $L$ is even ($L+1$ is odd), however, one has an additional symmetry requirement,
$n_j = n_{L-j}$, for both $\psi^{(+)}_0$ and $\psi^{(-)}_0$,
and the states differ in their electrons number:
$\psi^{(+)}_0$ contains $N=L/2$ electrons, for any $I>2$, 
while $\psi^{(-)}_0$ has $N=L/2+1$. 
It is worth noting that $n_1^{(+)} > 1/2$,
while $n_1^{(-)} < 1/2$.

\begin{figure}[htb]\centering
\vskip -2.0truecm
\epsfxsize6.0cm
\epsfbox{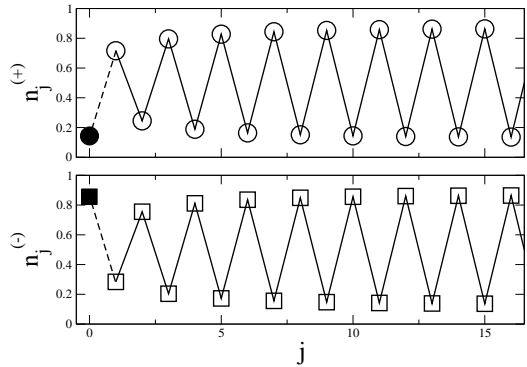}
\vskip -.5truecm
\caption{\label{fig:cdw_dot_scheme}
The occupation of the dot and the first lead sites of the
CDW states $\psi_0^{(+)}$ (circles)
and $\psi_0^{(-)}$
(squares), of a lead with 300 sites and $I=4$ which is 
coupled to the dot (filled symbols) with $V=0.8$.
For $\epsilon_0=0$ these two states are degenerate in the thermodynamic limit.
Lines are guide to the eye.
}
\end{figure}

\begin{figure}[htb]\centering
\vskip -2.0truecm
\epsfxsize6.0cm
\epsfbox{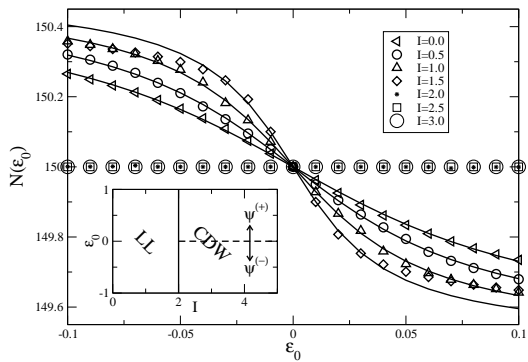}
\vskip -.5truecm
\caption{\label{fig:n_tot}
The total number of electrons in the system,
as a function of $\epsilon_0$, for $L=299$, $\mu=0$, $V=0.3$, and
different values of $I$. Symbols - DMRG results.
Lines - fit to Eq.~\ref{eqn:n_miu}.
Inset: the phase diagram: a Kosterlitz Thouless transition (solid line)
occurs at $I=2$, and a first order phase transition (dashed line) at 
$\epsilon_0=0$ for $I>2$.
}
\end{figure}

When $\epsilon_0 \ne 0$, i.e., a dot with a finite on-site energy is connected to the lead, it
has some local influence on the ground state in its vicinity. 
For a lead in the CDW phase, the two states 
described above are slightly modified, and can be denoted by $\psi^{(+)}$ 
and $\psi^{(-)}$. Nevertheless, the 
main influence of the dot is to lift the degeneracy between those states.
If $\epsilon_0<0$, the dot state population is high and $\psi^{(-)}$
is energetically preferable due to the dot-lead hopping term.
For $\epsilon_0>0$ the opposite happens, resulting
in a preference of $\psi^{(+)}$. 
The occupation of the dot-lead system in the CDW phase is shown 
in Fig.~\ref{fig:cdw_dot_scheme}.
When the dot's orbital energy changes from a negative
to a positive value, the system switches from $\psi^{(-)}$ to
$\psi^{(+)}$. In the following we show, by a
calculation of the free energy, that this is indeed a
first order QPT. The resulting phase diagram is drawn in the inset of 
Fig.~\ref{fig:n_tot}.

Before discussing the free energy in the CDW phase,
we turn to an exact calculation of the energy
and of the total number of electrons in a non-interacting case.
Using Green functions technique \cite{mahan}, one can find that
the change in the total density of states in the system,
due to the presence of the dot, is
\begin{eqnarray} \label{eqn:dos}
\Delta \nu(\epsilon) = \frac {1}{\pi} 
\Im \frac {\partial} {\partial \epsilon} \ln (\epsilon -\epsilon_{0} - \Sigma(\epsilon)),
\end{eqnarray}
where $\Sigma(\epsilon) = (V/t)^2\epsilon/2 + i (V^2/t) \sqrt{1-(\epsilon/2t)^2}$ 
is the self energy of an electron in the dot \cite {our_prev}.

Therefore, the change in the number of electrons in the entire 
system is
\begin{eqnarray} \label{eqn:n_miu}
\Delta N(\mu,\epsilon_0) = \int_{-2t}^{\mu} {\Delta \nu(\epsilon) d\epsilon} = \\ \nonumber
\frac {1}{2} + \frac {1}{\pi} & \displaystyle 
\tan^{-1}{ \frac {\mu -\epsilon_{0} - \Re \Sigma(\mu)} {\Im \Sigma(\mu)}},
\end{eqnarray}
and the change in the free energy of the system is
\begin{eqnarray} \label{eqn:free_E}
\Delta \Omega(\mu,\epsilon_0) = \Delta E - \mu \Delta N(\mu) = \\ \nonumber
\int_{-2t}^{\mu} (\epsilon-\mu) \Delta \nu(\epsilon) d\epsilon &=& 
- \int_{-2t}^{\mu} \Delta N(\epsilon,\epsilon_0) d\epsilon.
\end{eqnarray}

In Fig.~\ref{fig:n_tot} the total number of electrons 
in the system is presented. For $I=0$, 
$N$ fits the predicted formula (Eq.~\ref{eqn:n_miu}) well. 
As noted above, the agreement between the numerical results for a 
finite lead and the exact result for a semi-infinite lead, is obtained 
due to the fact that the number of particles can vary during the DMRG process.
For the LL phase ($0<I<2$),
$N(\epsilon_0)$ looks quite similar, varying between 
$L/2+1$ (at $\epsilon_{0} \rightarrow -\infty$)
and $L/2$ (at $\epsilon_{0} \rightarrow \infty$),
taking the average value $(L+1)/2$ at $\epsilon_{0}=0$. As in the case of
$n_{dot}$, the results fit Eq.~(\ref{eqn:n_miu}) with a
renormalized dot-lead coupling for small values of $I$. Increasing $I$ 
towards the transition point ($I > 1.5$), results in a less accurate fit.

The values of $V_{eff}(I)$ obtained by the fit of the $N$ curves 
to Eq.~(\ref{eqn:n_miu}), 
are in good agreement with the values obtained by the fit of $n_{dot}$ 
to Eq.~(\ref{eqn:n_dot_exact}). We find that $V_{eff}$ decreases
monotonically with increasing $I$, exhibiting the RPA like
behavior described in Ref.~\onlinecite{our_prev}.

The CDW phase ($I>2$), however,
is qualitatively different: for an odd lead length $N(\epsilon_0)$
remains a constant integer ($N=(L+1)/2$) which
does not depend on $\epsilon_0$ at all.
For even $L$, $N=L/2+1$ for $\epsilon_0<0$, and
$N=L/2$ for $\epsilon_0>0$.
This is a direct result from the 
switch of the system ground state from 
$\psi^{(-)}$ to $\psi^{(+)}$
at $\epsilon_0=0$.
For an odd lead length, the total number of
sites in the system ($L+1$) is even, so that $N$ is equal for both states. 
For an even lead length, $L+1$ is odd, and the total number of 
electrons is changed by one when $\epsilon_0$ passes $0$.
Except for the decrease of one electron at $\epsilon_0=0$ for an even lead,
$N$ remains constant in the CDW phase, independent of $\epsilon_0$.
Thus even with the continuous change in the population of the
dot as a function of the dot's level for $\epsilon_0 \ne 0$,
the number of electrons in the entire system remains constant. 
The change in the occupation of the dot as a function of $\epsilon_0$
is compensated by the lead.

This difference in the behavior of $N$ as a function of $\epsilon_0$
between the two phases (i.e., constant for
the CDW phase, compared to a continuous decrease 
for the LL phase) is a direct manifestation of their transport properties.
The LL phase is metallic, and therefore compressible. Hence, infinitesimal 
changes of the electrons number are possible. On the other hand, the CDW phase
is insulating and thus incompressible, which results in a constant $N$.

\begin{figure}[htb]\centering
\vskip -2.0truecm
\epsfxsize6.0cm
\epsfbox{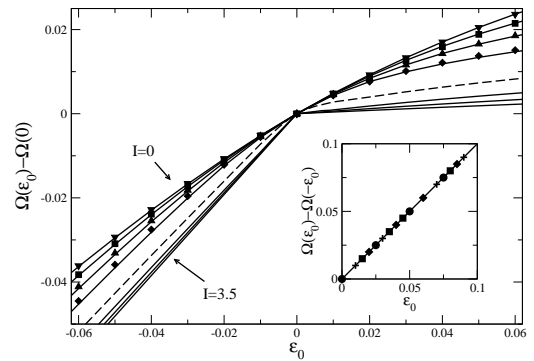}
\vskip -.5truecm
\caption{\label{fig:free_energy}
The ground state free energy as a function of $\epsilon_0$ for different 
interaction strengths. The lead-dot coupling was taken as
$V=0.3$, and the interaction strength
$I$ takes values between $0$ and $3.5$, in jumps of $0.5$.
The lines represent the DMRG results (dashed line - $I=2$), while the symbols show the 
fit to Eq.~\ref{eqn:free_E} for values of $I$ lower than $2$.
Inset: $\Omega(\epsilon_0)-\Omega(-\epsilon_0)$ 
as a function of $\epsilon_0$ for $I=0,1,2,3$ (symbols) 
and $\Omega(\epsilon_0)-\Omega(-\epsilon_0) = \epsilon_0$ (line).
For all values of interaction we get
$\left| \Omega(\epsilon_0)-\Omega(-\epsilon_0) - \epsilon_0 \right| < 10^{-6}$.}
\end{figure}

In Fig.~\ref{fig:free_energy}, typical numerical results for 
the free energy $\Omega(\epsilon_0)-\Omega(0)$ as 
a function of $\epsilon_0$ are shown. 
A perfect fit between the DMRG results
and the exact formula Eq.~(\ref{eqn:free_E}) for $I=0$ is obtained.
In the LL phase ($0<I<2$) our DMRG calculations show 
that the effect of interactions on $\Omega$ can be fitted by
replacing $V$ in Eq.~(\ref{eqn:free_E}) by the same
effective coupling $V_{eff}$ obtained for the behavior
of $n_{dot}$ discussed in Ref.~\onlinecite{our_prev},
and for the behavior of $N$ discussed above. 
For the CDW phase ($I>2$), however,
there is obviously a qualitative change in the energy curve:
the dependence of $\Omega$ on $\epsilon_0$ is linear
both below and above $\epsilon_0=0$, with an abrupt
change of $\frac {d\Omega}{d\epsilon_0}$ at $\epsilon_{0}=0$.

These results point out that the single impurity,
connected at one end of a long interacting lead, has a well defined
influence on the ground state of the entire coupled system. As
discussed above, when the dot level passes through $\epsilon_0=0$,
the lead's population is inverted at every site, leading to
an inversion of the CDW order parameter, 
as presented in Fig.~\ref{fig:cdw_order}(inset). 
The dot population is inverted as well. As a result,
a dramatic change in the dependence of the free energy of the system
on $\epsilon_{0}$ occurs.

\begin{figure}[htb]\centering
\vskip 0.6truecm
\epsfxsize8.0cm
\epsfbox{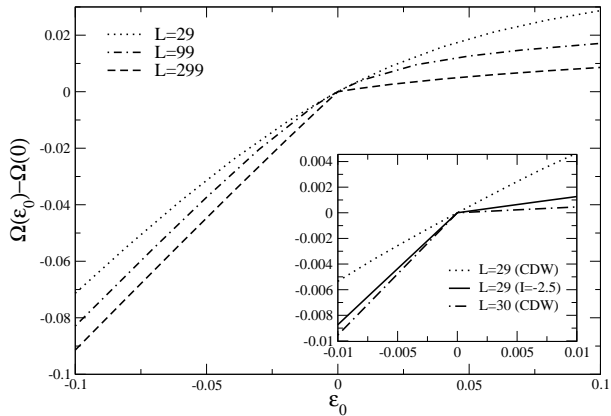}
\caption{\label{fig:qpt_L}
The ground state free energy as a function of $\epsilon_0$ for different 
CDW system sizes, with $V=0.3$ and $I=2.5$ (dotted, dashed-dotted and dashed lines).
Inset: the results for $L=29$ are compared to the case of $I=-2.5$, 
which is inside the ferromagnetic phase, and to the CDW case with $L=30$.
}
\end{figure}

Since $\Omega$ is the free energy of the system,
the jump in it's first derivative might be a sign of a first order QPT. 
In order to see whether this non-analyticity of the free
energy is just a trivial level crossing (LC) of two levels in the system, 
the dependence of the transition shape on the system size $L$ is explored.
For the case that the sharp transition in the energy results from the fact
that the external field (in our case the gate voltage) commutes with the
Hamiltonian, and thus a LC is possible, no size dependence of the sharpness 
of the transition is expected. On the other hand, a real QPT will become 
sharp only in the thermodynamic limit (i.e., semi infinite lead).

In order to compare the two scenarios (LC vs. QPT) we solve
the same Hamiltonian (Eq. (1)) with strong
attractive interactions, i.e. $I<-2t$. In this case, transforming
the Hamiltonian to an XXZ spin chain, the strong attractive interactions
regime corresponds to the ferromagnetic
phase \cite{kosterlitz} (while the CDW is equivalent to the N\'eel phase).
In the fermionic case these attractive interactions yield a two-fold 
degenerate ground state, composed of states in which the sites
are either entirely occupied or 
empty (in case there is an explicit restriction to half filling, 
a "phase separation" evolves).
Pinning by the dot lifts the degeneracy, because an occupied dot
causes a preference of an empty lead, and vice versa.
Thus the influence of the dot on the coupled wire is superficially similar 
to the CDW case in so far as in both cases the ground state 
of the entire system is determined by the dot orbital. 
Nevertheless, the ferromagnetic system is 
clearly a LC system, and no dependence on the system size is expected.

The dependence of $\Omega$ on $\epsilon_0$ was calculated for 
$L=29,99,299$ in the CDW case, and for $L=29$ in the ferromagnetic case.
From the results shown in Fig.~\ref{fig:qpt_L} (and its inset)
it is clear that these two cases are different.
The ferromagnetic system is indeed a trivial LC system, with two
competing states whose energies cross each other for $\epsilon_0=0$.
Since these two states are eigenstates of the Hamiltonian even for
a finite $L$, the size does not play a role, so that even for $L=29$
one can see a sharp transition between the two ground states.

On the other hand, for a small (but with an odd $L$) CDW system $\Omega$ 
shows a smooth dependence on $\epsilon_0$, without any non-analyticity. 
As a matter of fact, for any finite system the electron levels are expected
to be mixed, resulting in avoided crossing of the two lowest many-body 
levels \cite{qpt_book}. In other words, the CDW states are not true 
eigenstates of the Hamiltonian for a finite system. Indeed,
indications for the sudden jump of $\frac {d\Omega}{d\epsilon_0}$ are seen 
only for larger system sizes (i.e., for $L \gtrsim 200$).

The dot-dashed line in Fig.~\ref{fig:qpt_L}(inset) shows that for a short lead 
of an even size ($L=30$ in that case) there is, however, a non-analyticity 
in $\Omega$ as a function of $\epsilon_0$. The comparison between even and odd 
lead sizes emphasizes our conclusion stated above. 
While for an odd lead size $\psi^{(-)}$ and $\psi^{(+)}$ have the
same number of electrons, for an even size of the lead, the transition 
from $\psi^{(-)}$ to $\psi^{(+)}$ at $\epsilon_0=0$ involves a decrease 
of the electrons number by one. 
The Hamiltonian $\hat H$ conserves the number of particles, so that in
the case of $L$ even, these two states are not coupled, and the 
transition between them is a simple LC, not showing a size dependence. 
For an odd $L$, however, $\psi^{(-)}$ and $\psi^{(+)}$ are coupled
by $\hat H$, so that for a finite $L$ they are actually mixtures of the 
CDW states, thus presenting sharper $\Omega(\epsilon_0)$ dependence 
for larger systems. 
We thus conclude, based on these two comparisons, that in the case of a CDW 
with an odd $L$ this transition is a QPT, which happens for 
$L \rightarrow \infty$, i.e., for a semi-infinite lead\cite{QPT_note}.

\begin{figure}[htb]\centering
\vskip 0.6truecm
\epsfxsize8.0cm
\epsfbox{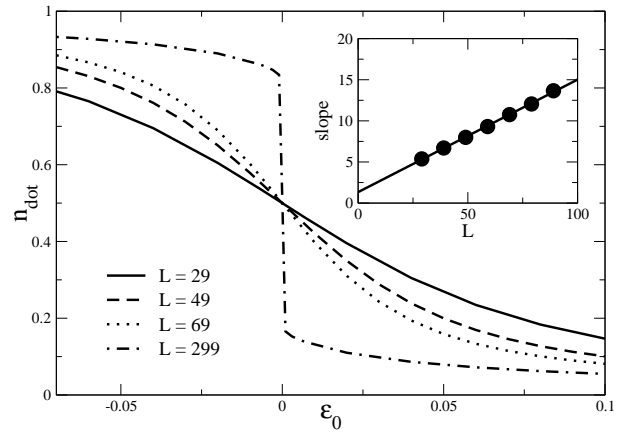}
\caption{\label{fig:scaling_L}
The dependence of the dot occupation on $\epsilon_0$ for different CDW system sizes, 
with $V=0.3$ and $I=2.5$.
Inset: the absolute value of the slope in the limit $\epsilon_0 \rightarrow 0$ 
as a function of $L$ for short leads, together with a linear fit.
}
\end{figure}

As pointed out above, in the numerical results for large lead sizes, the
non-analyticity of the free energy is seen clearly, so that it is obviously
a first order transition. However, a similar conclusion can be drawn from
the results of short leads, by scaling the results against the lead
size $L$. 
Although for finite size systems the order parameter changes continuously,
the slope of this change grows with the system size $L$.
For a first order transition in $d$ dimensions one expects a power law dependence 
of the slope as $L^d$, while for a second order transition
the power law should be fixed by some universal exponents \cite{binder84}.

For the CDW model, the population of the dot plays the role 
of an order parameter, and one can check its behavior near $\epsilon_0=0$ for different 
lead sizes. In Fig.~\ref{fig:scaling_L} the occupation as a function of $\epsilon_0$ 
is shown for some short lead sizes ($L=29,49,69$) and for a large one ($L=299$). 
It is indeed seen that while for a long enough lead there is a jump in the
dot population, for short leads there is a continuous change in the occupation, 
with a larger slope for larger lengths. Zooming into the regime of $\epsilon_0=0$
we find that although $L \gtrsim 200$ is required in order to see a clear discontinuity
in $\Omega$,
$L \gtrsim 100$ is long enough for showing a jump in $n_{dot}$.
The dependence of the slope on $L$, for $L<100$,
is shown in the inset of Fig.~\ref{fig:scaling_L}. As one expects from
finite size scaling predictions for first order phase transition,
there is a very good linear fit of the the order parameter on the lead size.

We now turn to another consequence of the particle-hole symmetry.
For $\epsilon_0 \ne 0$ the symmetry results in the fact that the
ground states for $+\epsilon_0$ and for $-\epsilon_0$ 
($\psi^{(+)}$ and $\psi^{(-)}$ respectively)
have an inverted population everywhere.
It is thus clear that $\langle \psi^{(+)}|\hat H_{lead}|\psi^{(+)} \rangle = 
\langle \psi^{(-)}|\hat H_{lead}|\psi^{(-)} \rangle$, and similarly
$\langle \psi^{(+)}|\hat H_{dot-lead}|\psi^{(+)} \rangle = 
\langle \psi^{(-)}|\hat H_{dot-lead}|\psi^{(-)} \rangle$. This implies
\begin{eqnarray} \label{eqn:E_diff}
\Omega(\epsilon_0)-\Omega(-\epsilon_0) &=& \\ \nonumber
\langle \psi^{(+)}|\hat H_{dot}|\psi^{(+)} \rangle &-&
\langle \psi^{(-)}|\hat H_{dot}|\psi^{(-)} \rangle = \\ \nonumber
&& \epsilon_0 n_{dot}^+ - (-\epsilon_0) n_{dot}^- = \epsilon_0,
\end{eqnarray}
where $n_{dot}^+$ ($n_{dot}^-$) represents the dot populations of 
$\psi^{(+)}$ ($\psi^{(-)}$).
The last equality results from the symmetric population
$n_{dot}^- = 1 - n_{dot}^+$.

The relation Eq.~\ref{eqn:E_diff} does not depend on the interaction,
and should exist for both the LL and the CDW phases. 
Obviously it should be obeyed for the non-interacting case,
where the energy is given by Eq.~(\ref{eqn:free_E}), thus
leading to the following nontrivial relation
\begin{eqnarray} \label{eqn:E_diff_I0}
{ 
- \frac {1}{\pi} \int_{-2t}^{2t}
\tan^{-1} \frac {  (1- \frac {V^2} {2t^2}) \epsilon -\epsilon_{0} }
{ \frac {V^2}{t} \sqrt {1- \frac {\epsilon^2}{4t^2}} }
 d\epsilon
= \epsilon_{0}},
\end{eqnarray}
in which the result of the integral on the LHS does not depend on the parameters 
$V$ and $t$. 

A physical insight into Eq.~(\ref{eqn:E_diff_I0}) can be gained by
taking the derivative of both sides with respect to $\epsilon_0$, and using the
definition of the self energy given above. This yields
\begin{eqnarray} \label{eqn:E_tag_diff}
{\frac {1}{\pi} \int_{-2t}^{2t}
\frac {\Im \Sigma(\epsilon)} {(\epsilon -\epsilon_{0} - \Re \Sigma(\epsilon))^2 + 
(\Im \Sigma(\epsilon))^2} d\epsilon = 1},
\end{eqnarray}
which is evident since the LHS is 
the occupation of the dot \cite{our_prev} when $\mu > 2t$.

In Fig.~\ref{fig:free_energy}(inset) a plot of 
$\Omega(\epsilon_0)-\Omega(-\epsilon_0)$ 
as a function of $\epsilon_0$ is shown, for values of $I$ between $0$ and $3$.
As can be seen, Eq.(\ref{eqn:E_diff}) is valid 
for all values of $I$. 

In conclusion, we have shown that the occupation of a dot coupled to a
one dimensional wire can be used to identify the different phases in
the wire. It is known that as a function of the interaction strength $I$ 
the wire goes through three different phases (which may also be mapped 
onto the phases of an XXZ spin chain).
For $I<-2t$ the ground state is doubly degenerate, with all sites 
equally full or empty, corresponding to the ferromagnetic state of a spin chain.
In the intermediate range of 
interactions ($-2t<I<2t$) the translational symmetry is unbroken 
and a Luttinger Liquid phase occurs. For $I>2t$ 
the ground state is doubly degenerate 
corresponding to the two possible CDW states or the two different
antiferromagnetic realizations for the spin chain.

For the Luttinger liquid phase the occupation of the impurity does not 
show a jump when the impurity level crosses the Fermi energy. 
In the two other phases the impurity level splits the 
degeneracy by favoring one of the two ground states depending on whether the 
impurity level is empty or filled. 
Nevertheless, we have shown that the physics in these cases is different.
While for the ferromagnetic phase a simple level crossing occurs
with a sharp jump in occupation of the impurity for any length of the wire,
in the CDW phase the position of the impurity level drives 
a first order QPT in the thermodynamical limit between the two CDW 
states, while for a finite wire the jump is smeared. This phase transition
shows all the hallmarks of a first order QPT, such as a size dependence, 
a jump in the order parameter and a discontinuity of the derivative of the 
grand canonical potential. 

\acknowledgments

Support from the Israel Academy of Science (Grant 877/04) is gratefully
acknowledged.


\begin{thebibliography}{2}
\bibitem{QPT_note} Of course, when $\epsilon_0$ crosses the Fermi energy, the system will
remain in the previous ground state, which becomes a metastable state. In the thermodynamic 
limit an exponentially long time will be required for the system to switch to the true
ground state.
\end{thebibliography}
\end{document}